\documentclass[11pt]{revtex4-1}
\usepackage{bm}
\usepackage{color,graphicx}
\usepackage{tensor}
\usepackage{amsmath}
\begin{document}
\newcommand{\boldnabla}{\bm{\nabla}}
\newcommand{\boldtheta}{\bm{\theta}}
\newcommand{\angmom}{\bm{\mathcal{J}}}


\title{Optical angular momentum in atomic transitions: a paradox}

\author{Stephen M.~Barnett}
\email{stephen.barnett@glasgow.ac.uk}
\author{Fiona C.~Speirits}
\email{fiona.speirits@glasgow.ac.uk}
\address{School of Physics and Astronomy, University of Glasgow,
Glasgow G12 8QQ, United Kingdom}
\author{Mohamed Babiker}
\address{Department of Physics, University of York, Heslington, York, YO10 5DD, United Kingdom}

\date{\today}


\begin{abstract}
\noindent Stated simply the paradox is as follows: it is clear that the
orbital angular momentum of a light beam in its direction of propagation is
an intrinsic quantity, and therefore has the same value everywhere in the
beam.  How then can a Gaussian beam, with precisely zero orbital angular
momentum, drive a (single-photon) quadrupole transition which requires the
transfer of angular momentum $2\hbar$ to an absorbing atom?
\end{abstract}


\maketitle



\section{Introduction}

It has long been appreciated that light has mechanical properties including
energy, momentum and also angular momentum.  The energy property was known to
the ancients but the first suggestion of the momentum appears to be by Kepler
who observed that the dust tails of comets point away from the sun and ascribed
this effect to radiation pressure \cite{Kepler}.  It was Maxwell who first
calculated the radiation pressure due to sunlight on the surface of the earth
\cite{Maxwell} and Poynting who quantified the momentum and energy flux
associated with the electromagnetic field \cite{Poyntingvec,Poyntingbook}.
Poynting also proposed that light can carry angular momentum and suggested, in
particular, that angular momentum was associated with circular polarisation
\cite{Poyntingspin}.

The modern study of optical angular momentum, and in particular of orbital
angular momentum, can be traced to the work of Allen {\it et al} \cite{Les} who
showed that laser modes, specifically the Laguerre-Gaussian modes, carry
$\ell\hbar$ units of orbital angular momentum about the beam axis for each
photon, where $\ell$ is the charge of the phase vortex at the centre of the
mode.  Within the paraxial regime we can assign also a separate spin part of
$\pm\hbar$ per photon associated with the circular polarisation.  The
combination of these orbital and spin parts of the optical angular momentum has
proven to be most versatile and has found numerous and diverse applications
\cite{OAMbook,Bekshaev,Alison,Andrews,PhilTrans}.

Optical angular momentum plays a crucial role in the interaction between light
and matter.  Selection rules between atomic and molecular energy levels are
determined in terms of changes in angular momentum
\cite{Condon,Kuhn,CohenT,Sobelman}, as are transitions occurring in nuclei
\cite{Rose}.  The strongest transitions in an atom are those mediated by the
electric dipole interaction.  For these the conservation of angular momentum
imposes selection rules on the electronic total angular momentum and on the
angular momentum component quantum numbers in the form $\Delta l = \pm 1$ and
$\Delta m = 0,\pm 1$.  (It is an accident of history that the letter $l$ is
used both for the $z$-component of the optical angular momentum and also for
the electronic total angular momentum quantum number.  In the hope of avoiding
confusion we use different fonts for these two quantities, $\ell$ for the
optical orbital angular momentum and $l$ for the total electronic angular
momentum quantum number.)  The required angular momentum is readily provided by
the absorption of a photon with the appropriate polarisation; in particular the
absorption of a left- or right-circularly polarised photon exciting a
transition with $\Delta m = +1$ or $-1$ respectively.  The absorption of a
photon in this way transfers energy to excite the atom and spin angular
momentum to change the internal angular momentum.  It is not possible for the
required angular momentum to be provided by the orbital angular momentum of the
field \cite{Bennett}.  The absorbed photon transfers, also, linear momentum and
also orbital angular momentum to the centre of mass motion of the atom
\cite{Lembessis}.  The last of these has its strongest and most subtle effect
when the atom is in the vicinity of an optical vortex
\cite{BerryDennis,Qcore,superkick}.

In this paper we are principally concerned with angular momentum transfer to an
atom via higher-order multipolar transitions and, in particular, by an electric
quadrupole transition, in which the angular momentum of the atom changes by
$\Delta l = +2$, $\Delta m = +2$.  We have in mind something like the
$5^2S_{1/2}$ to $5^2D_{5/2}$ transition in Rubidium 87 \cite{Kien}, but our
analysis is not restricted to a particular transition or atom.  In an elegant
experiment, Afanasev {\it et al} showed that such a transition can be enhanced
if the atom (in the experiment it was a calcium 40 ion) is held at the vortex
core of a suitably prepared beam.  In one case in particular, a circularly
polarised $\ell = 1$ beam was employed, so that the required two quanta of
angular momentum are provided by one of optical spin angular momentum and one
of optical orbital angular momentum \cite{Afanasev}.

\begin{figure}[h!]
\centering
\includegraphics[width=0.3\textwidth]{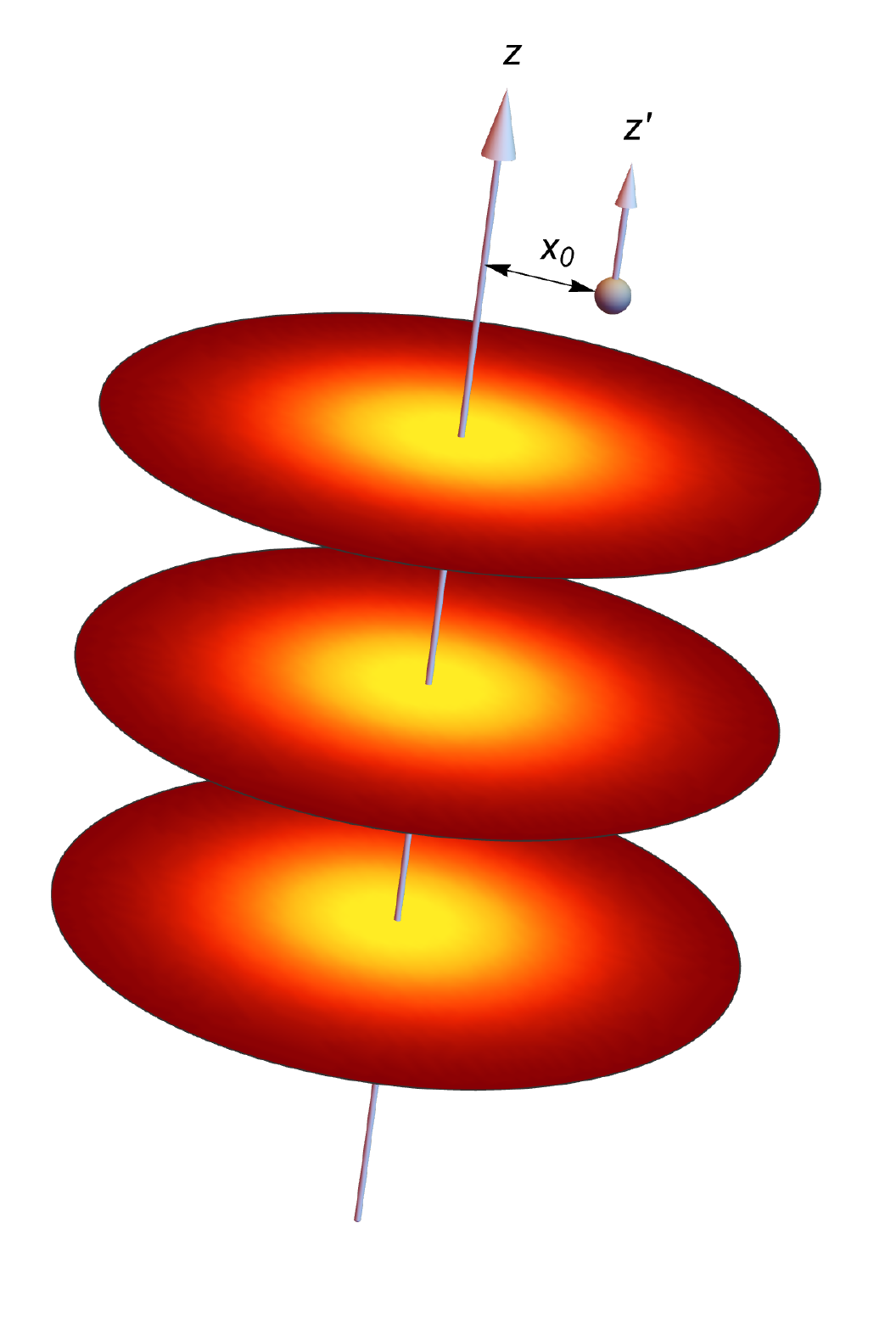}
\caption{An atom displaced a distance $x_0$ from the z-axis of a Gaussian beam.
This beam has flat phase fronts and therefore the orbital angular momentum about
the beam's $z$-axis is zero. The orbital angular momentum about the atom's axis $z^{\prime}$
is also identically zero, as orbital angular momentum is intrinsic, and therefore we must ask
from where the atom can obtain
the two quanta of angular momenta required to drive the stated quadrupole transition.}
\label{gaussian_beam}
\end{figure}

The paradox in our title arises when we consider driving the quadrupole
transition by a circularly polarised beam of light carrying zero units of
orbital angular momentum, as shown in Figure~\ref{gaussian_beam}.  The phase fronts are
approximately perpendicular to the propagation direction and do not have the helical form
characteristic of the presence of orbital angular momentum \cite{Les,OAMbook,Bekshaev,Alison,Andrews,PhilTrans}.  Such a beam
carries only one quantum of spin angular momentum per photon and, moreover, as
{\it both} the spin and the orbital angular momenta are intrinsic, it appears
to be the case that whatever the position of the atom, no quanta of orbital
angular momenta are available. How then is the angular momentum conserved in
such a transition?


\section{Optical orbital angular momentum: intrinsic or extrinsic?}

In mechanics it is often helpful and straightforward to separate spin and
orbital angular momentum.  The motion of the Earth, for example, is readily
separated into an intrinsic spin component (responsible for the cycle of days
and nights) and an extrinsic orbital motion associated with the calendar year.
The former is intrinsic in that its value does not depend on the position of
the chosen axis, but the latter does depend on the choice of axis and so is
extrinsic.  For light the situation is more subtle.  We have long known how to
extract spin and orbital parts \cite{Darwin} but it is clear that, although
these are separately measurable, neither is a true angular momentum
\cite{vanEnk,Rot,natures}.

We are concerned with the component of the optical angular momentum in the
direction of propagation of our light beam.  Here we can readily identify
separate spin and orbital parts of the total angular momentum and, indeed,
observe the individual contributions from these \cite{Anna}. However in this
case \emph{both} the spin and orbital parts are intrinsic, that is they are
unchanged if we evaluate them about any axis parallel to the direction of the
beam.  Berry provided a simple and general proof of this, which we reproduce
here \cite{Berry}.

The spin part of the optical angular momentum is intrinsic and we can determine
the extrinsic or intrinsic nature of the orbital part by examining the total
angular momentum.  Consider a beam of light propagating in the $z$-direction,
carrying some quantity of the $z$-component of angular momentum.  If we shift by ${\bf r}_0$ the
axis about which the angular momentum is evaluated, then we find that the total
angular momentum changes to
\begin{equation}
\label{Eq1}
J'_z = J_z - ({\bf r}_0\times{\bf p})_z \, ,
\end{equation}
where ${\bf p}$ is the total momentum of the beam.  Yet if the beam is
propagating in the $z$-direction, then the transverse components of the total
momentum of the light $p_x$ and $p_y$, are both zero and we conclude that the
angular momentum of the beam is independent of the position about which it is
determined: $J'_z = J_z$.  The spin angular momentum is intrinsic and it
follows, therefore, that the orbital part is also intrinsic.


\section{Electric dipole and electric quadrupole transitions}

To highlight the issue of angular momentum conservation, we present here simple
analyses of a $\Delta m = +1$ electric dipole transition and a $\Delta m = +2$
quadrupole transition, both driven by a circularly polarised Gaussian beam of
light which carries $\hbar$ units of spin angular momentum per photon, but zero
units of orbital angular momentum.  For definiteness we consider a pair of
transitions from the ground state of atomic Rubidium 87, $5^2S_{1/2}$, to the
$5^2P_{3/2}$ level and to the $4^2D_{5/2}$ level.  The first of these is an
electric dipole transition with $\Delta l = +1$ and $\Delta m = +1$,
corresponding to the gain, by the atom, of one quantum $\hbar$ of angular
momentum.  The second is an electric quadrupole transition in which the atom
acquires two quanta, $2\hbar$, of angular momentum from the single photon
absorbed.

\subsection{Electric dipole transition}

The electric dipole transition is associated with the interaction energy of the
form \cite{Edwin,Rodney}
\begin{equation}
\label{Eq2}
V_d = -D_iE_i \, ,
\end{equation}
where ${\bf D}$ is the dipole moment operator for the atom and ${\bf E}$ is the
resonant driving electric field.  Here we adopt the summation convention in
which repeated indices imply a summation over the three cartesian coordinates.
The dipole moment has the form
\begin{equation}
\label{Eq3}
D_i = -e\sum_\alpha r^\alpha_i
\end{equation}
where the summation runs over all the atomic charges.  For Rubidium we need
consider only the valence electron and so replace the summation by the
coordinates (relative to the nucleus) of this single electron, so the dipole
moment operator becomes
\begin{equation}
\label{Eq4}
D_i = -e r_i \, .
\end{equation}

Let us consider the interaction matrix element for the transition from the
ground state, $5^2S_{1/2}$, to the excited state, $5^2P_{3/2}$. As we are
increasing the energy of the atom, we can associate this with photon absorption
and therefore with the positive frequency, $e^{-i\omega t}$, part of the
electric field.  We consider a monochromatic paraxial (or weakly focussed) beam
of light propagating in the $z$-direction.  Hence we can write the transition
matrix element from the ground state ($|g\rangle = |5, 0, 0, 1/2\rangle$,
where the four entries in this state correspond to the quantum numbers $n$,
$l$, $m$ and $s$) to the excited state ($|e\rangle = |5,1,1,1/2\rangle$) in the form
\begin{equation}
\label{Eq5}
\langle e|V_d |g\rangle = e\langle e|r_i|g\rangle E_i({\bf R}) e^{-i\omega t} \, .
\end{equation}
Here $E_i({\bf R})$ is the space-dependent part of the complex electric field
at the position of the atom.  The dominant components of the electric field lie
in the $x - y$ plane and there is no need to consider, for our purposes, the
much smaller component in the $z$-direction.

To determine the optimal form of the field with which to drive the transition,
we need to evaluate the required form of the dipole matrix element. To this end
we note that the wavefunction for the electron to which the field is coupled
can be written in the form
\begin{equation}
\label{Eq6}
\psi_{n,l,m}({\bf r}) = R_{n,l}Y^m_l(\vartheta,\varphi) \, ,
\end{equation}
where $Y^m_l(\vartheta,\varphi)$ is the spherical harmonic
\cite{Condon,Kuhn,CohenT,Sobelman} and we have omitted mention of the electron
spin, which does not change in the transition.  We then find that the dipole
matrix element is
\begin{align}
\label{Eq7}
d_i &= \langle e|D_i|g\rangle \nonumber \\
&= -e\int_0^\infty r^2dr\int_0^\pi \sin\vartheta\int_0^{2\pi}d\varphi R_{5,1}(r)Y^{1*}_1(\vartheta,\varphi)r_i
R_{5,0}(r)Y^0_0(\vartheta,\varphi) \, ,
\end{align}
where the spherical harmonics are
\begin{align}
\label{Eq8}
Y^0_0(\vartheta,\varphi) &= \frac{1}{2}\sqrt{\frac{1}{\pi}} \nonumber \\
Y^1_1(\vartheta,\varphi) &= -\sqrt{\frac{3}{8\pi}}\sin\vartheta e^{i\varphi} \, .
\end{align}
It follows that the components of the dipole matrix element are
\begin{align}
\label{Eq9}
d_x &= \frac{e}{\sqrt{6}}\int_0^\infty r^2dr R_{5,1}rR_{5,0} = d \nonumber \\
d_y &= -id \, .
\end{align}
Hence the matrix element of the interaction energy is
\begin{equation}
\label{Eq10}
\langle e|V_d|g\rangle = -d(E_x({\bf R})-iE_y({\bf R}))e^{-i\omega t} \, .
\end{equation}
To maximise the magnitude of this, and thereby find the maximum rate for
driving this transition, we simply set $E_y = iE_x$, so that the electric field
takes the form
\begin{equation}
\label{Eq11}
{\bf E}({\bf R}) = \frac{E_0({\bf R})}{\sqrt{2}}(\hat{\bf x} + i\hat{\bf y})e^{-i\omega t} \, ,
\end{equation}
where $\hat{\bf x}$ and $\hat{\bf y}$ are unit vectors in the $x$- and
$y$-directions. Note that this corresponds to a field with left-handed circular
polarisation, corresponding to $+\hbar$ spin angular momentum per photon.

Whatever the position of the absorbing atom, the intrinsic spin angular
momentum of the light provides the required single quantum of angular momentum
about the displaced $z$-axis centred on the position of the atom. The only
dependence on position comes from the spatial variation of the strength of the
field $E_0$ and hence the intensity.

\subsection{Electric quadrupole excitation}

The electric quadrupole interaction energy depends on the derivatives of the
electric field at the position of the atom:
\begin{equation}
\label{Eq12}
V_q = Q_{ij}\nabla_iE_j \, ,
\end{equation}
where $Q_{ij}$ is the quadrupole moment defined to be \cite{Edwin,Rodney}
\begin{equation}
\label{Eq13}
Q_{ij} = -\frac{1}{2}r_ir_j  \, .
\end{equation}
Note that both the electric dipole and electric quadrupole interactions can
readily be derived within the Power-Zienau-Woolley multipolar expansion
\cite{Bennett,Edwin}. We are interested in transitions in our Rubidium atom
from the ground state, $5^2P_{1/2}$, to the excited state, $4^2D_{5/2}$.  We
can write the transition matrix element from the ground to the excited state
($|e\rangle = |4,2,2,1/2\rangle$) in the form
\begin{equation}
\label{Eq14}
\langle e|V_q|\rangle = -\frac{1}{2}\langle e|r_ir_j|g\rangle \nabla_j E_i e^{-i\omega t} \, ,
\end{equation}
where the derivatives of the field are evaluated at the position of the atom
${\bf R}$.

To determine the form of the field required to drive this transition, we need
to evaluate the quadrupole matrix element. In order to calculate this we
require the spherical harmonic associated with the excited state:
\begin{equation}
\label{Eq15}
Y^2_2(\vartheta,\varphi) = \frac{1}{4}\sqrt{\frac{15}{2\pi}}\sin^2\vartheta e^{i2\varphi} \, .
\end{equation}
Hence our quadrupole matrix elements $\langle e|Q_{ij}|g\rangle$ are
\begin{align}
\label{Eq16}
q_{xx} &= \langle e|Q_{xx}|g\rangle  \nonumber \\
&= -\frac{e}{2}\frac{1}{\sqrt{30}}\int_0^\infty r^2dr R_{4,2}(r) r^2 R_{5,0}(r) \nonumber \\
&= \mathcal{Q} \nonumber \\
q_{xy} &= -i\mathcal{Q} = q_{yx} \nonumber \\
q_{yy} &= -\mathcal{Q} \, .
\end{align}

To proceed let us consider the possible forms for the driving field.  As noted
above, the dominant components will be in the $x$- and $y$-directions and it is
the derivatives of the field that couple to the atom.  To this end we expand
the components of the electric field as a Taylor-Maclaurin series in the form
\begin{align}
\label{Eq17}
E_x({\bf r}) &= E_x({\bf R}) e^{-i\omega t} +(x-X)\frac{\partial}{\partial x}E_x({\bf R})e^{-i\omega t}
+ (y-Y)\frac{\partial}{\partial y}E_x({\bf R})e^{-i\omega t}  + \cdots  \nonumber \\
&=E_x({\bf R}) e^{-i\omega t} + [\alpha (x-X) + \beta (y-Y)]E^{(1)}({\bf R})e^{-i\omega t} + \cdots \nonumber \\
E_y({\bf r}) &= E_y({\bf R}) e^{-i\omega t} +(x-X)\frac{\partial}{\partial x}E_y({\bf R})e^{-i\omega t}
+ (y-Y)\frac{\partial}{\partial y}E_y({\bf R})e^{-i\omega t}  + \cdots  \nonumber \\
&= E_y({\bf R}) e^{-i\omega t} + [\gamma (x-X) + \delta (y-Y)]E^{(1)}({\bf R})e^{-i\omega t} + \cdots \, .
\end{align}
Without loss of generality, we can fix select our four complex constants such
that $|\alpha|^2 + |\beta|^2 + |\gamma|^2 + |\delta|^2 = 2$.  With this choice
we fix $E^{(1)}({\bf R})$ to be
\begin{equation}
\label{Eq17a}
E^{(1)}({\bf R}) = \left. \frac{1}{\sqrt{2}}\left[\sum_{i,j = x,y}\left(\nabla_iE_j\right)\left(\nabla_iE^*_j\right)\right]^{1/2}\right|_{\bf R} \, .
\end{equation}
We can compare this with the required form for our particular transition, for
which we have calculated the quadrupole matrix elements, Eq.~(\ref{Eq16}).
Thus,
\begin{equation}
\label{Eq18}
Q_{ij}\nabla_iE_j = E^{(1)}e^{-i\omega t}\mathcal{Q}(\alpha -i\beta - i\gamma - \delta) \, .
\end{equation}
Clearly this is maximised if the magnitudes of each of our parameters
$\alpha,\beta, \gamma$ and $\delta$ are equal. There is one arbitrary phase,
but if we choose $\alpha$ to be real and positive, then the strongest driving
occurs for
\begin{equation}
\label{Eq19}
\alpha = \frac{1}{\sqrt{2}} \, , \quad  \beta = \frac{i}{\sqrt{2}} = \gamma \, , \quad \delta = -\frac{1}{\sqrt{2}} \, .
\end{equation}
Note that the fact that $\delta = -\alpha $ is a consequence of the
transversality of the electric field $(\nabla_iE_i = 0)$. The values of
$\alpha, \beta, \gamma$ and $\delta$ correspond to an electric field in the
vicinity of the atom of the form
\begin{equation}
\label{Eq20}
{\bf E}({\bf r}) = {\bf E}({\bf R})e^{-i\omega t} + E^{(1)}e^{-i\omega t}[(x-X) + i(y-Y)]\frac{1}{\sqrt{2}}(\hat{\bf x} + i\hat{\bf y}) \, ,
\end{equation}
the second term of which corresponds to a circularly polarised $\ell = +1$
beam, with a charge $+1$ phase vortex centred on the atom, at $(X,Y)$,
corresponding to $+\hbar$ of orbital angular momentum per photon.


\section{The paradox}

We have determined the required form of the electric field in order to drive
our quadrupole transition; it should be circularly polarised, with an $\ell =
+1$ phase vortex centred on the position of the atom.  Let us see how a simple
circularly polarised Gaussian beam performs.  To this end consider an electric
field of the form
\begin{equation}
\label{Eq21}
{\bf E} = E_0\frac{1}{\sqrt{2}}(\hat{\bf x} + i\hat{\bf y})\frac{1}{\sqrt{\pi}w_0}\exp\left(-\frac{(x^2+y^2)}{2w_0^2}\right)e^{-i\omega t} \ .
\end{equation}
It is straightforward to calculate the quadrupole coupling matrix element for
an atom at position $(X,Y)$ in this field:
\begin{equation}
\label{Eq22}
\langle e|V_q|g\rangle = E_0\mathcal{Q}\sqrt{\frac{2}{\pi}}\frac{1}{w^3_0}(X + iY)\exp\left(-\frac{(X^2+Y^2)}{2w^2_0}\right) \, .
\end{equation}
This is non-zero everywhere apart from along the $z$-axis, corresponding to the
centre of the beam, for which $X=0=Y$.  The perturbative transition
probability, which is proportional to the squared modulus of this matrix
element, is greatest for an atom at a distance $w_0$ from the centre of the
beam, where the gradient of the magnitude of the field is greatest.

It follows that a circularly polarised Gaussian beam with only \emph{one}
quantum of angular momentum per photon can readily excite a quadrupole
transition requiring \emph{two} quanta of angular momentum.  Moreover, because
the orbital angular momentum is an intrinsic quantity, the orbital angular
momentum of the beam about an axis passing through the atom is also zero.  The
puzzle, therefore, is where can the required second quantum of angular momentum
have come from?


\section{An insight from quantum theory}

We recall that a mode of well-defined orbital angular momentum is an eigenmode
of the differential operator $-i\partial/\partial\phi$.  Indeed it was the
similarity between this property and the form of the quantum mechanical
operator for the $z$-component of the angular momentum $L_z =
-i\hbar\partial/\partial\phi$ that led to the conclusion that such eigenmodes
carry $\ell\hbar$ of orbital angular momentum for each photon \cite{Les}.  The
$z$-component of the orbital angular momentum about an axis shifted from the
$z$-axis to $x = x_0$, $L'_z$ is related to that about the $z$-axis $L_z$ by
\begin{equation}
\label{Eq23}
L'_z = L_z -x_0p_y \, ,
\end{equation}
where $p_y$ is the operator corresponding to the $y$-component of the linear
momentum. As in Berry's demonstration, this means that the total values of
$L_z$ and  $L'_z$, which correspond to the expectation values in quantum
theory, are the same \cite{Berry}.  The operators $L_z$ and $L'_z$, however,
are incompatible in that they do not commute:
\begin{equation}
\label{Eq24}
[L'_z,L_z] = i\hbar x_0 p_x
\end{equation}
and it follows that $L_z$ and $L'_z$ do not share common eigenmodes.  Hence an
eigenmode of $\hat{L}_z$ must correspond to a \emph{superpositon} of eigenmodes
of $L'_z$.  It is worth remarking that this simple idea does not seem to appear
in texts concerned with the quantum theory of angular momentum
\cite{RoseAM,Brink,Edmonds,BiedenharnEd,BiedenharnBook,Zare}.  We note,
however, that the angular momenta about different axes have been shown to occur
for free electrons moving in a uniform magnetic field making it possible to
separate angular momenta associated with the cyclotron and diamagnetic angular
momenta \cite{Greenshields}.  It is worth noting that in the commutator
(\ref{Eq24}), $x_0$ can be arbitrarily large.  This is a reflection of the fact
that the difference between $L_z$ and $L'_z$ is the product of the
$y$-component of the momentum operator and $x_0$, the distance between the two
rotation axes considered.  This is reminiscent of the parallel axes theorem in
mechanics \cite{Kibble}.

Consider a normalised wavefunction of the same form as our Gaussian mode:
\begin{equation}
\label{Eq25}
\psi(x,y) = \frac{1}{\sqrt{\pi}w_0}\exp\left(-\frac{(x^2+y^2)}{2w^2_0}\right) \, .
\end{equation}
This is an eigenfunction of $L_z$ with eigenvalue $0$ and therefore the
expectation value of $L'_z$ is also $0$.  It is not an eigenfunction of $L'_z$,
however, and has the non-zero variance
\begin{equation}
\label{Eq26}
\Delta L'^2_z = x^2_0\langle p^2_y\rangle = \frac{\hbar^2x^2_0}{2w^2_0} \, .
\end{equation}
When expressed in terms of the corresponding orbital angular momentum numbers
$\ell$ and $\ell'$ (associated with the operators $L_z$ and $L'_z$
respectively) this variance becomes
\begin{equation}
\label{Eq27}
\Delta\ell'^2 = \frac{x^2_0}{2w^2_0} \, .
\end{equation}
This indicates that our mode with zero orbital angular momentum about the
$z$-axis includes a superposition of a range of eigenmodes of the orbital
angular momentum about our displaced axis and suggests that our atom,
undergoing a quadrupole transition, absorbs a photon from the $\ell' = +1$
eigenmodes of $L'_z$ within this superposition.

To test the idea suggested above we seek to expand our Gaussian mode in terms
of a complete set of angular-momentum eigenmodes centred on the position of the
atom.  We introduce a primed set of coordinates, $(x',y')$ or $(\rho',\phi')$,
centred on the position of the atom.  In terms of these coordinates our
normalised Gaussian mode has the form
\begin{equation}
\label{Eq28}
\psi(x',y') = \frac{1}{\sqrt{\pi}w_0}\exp\left(-\frac{[(x'+x_0)^2+y'^2]}{2w^2_0}\right) \, .
\end{equation}
The Laguerre-Gaussian modes are all eigenmodes of $L'_z$ angular momentum and
we can use them as a complete basis \cite{Siegman,Milonni}:
\begin{equation}
\label{Eq29}
u_{p'\ell'}(\rho',\phi') = \frac{1}{w_0}\left(\frac{p'!}{\pi(|\ell'| + p')!}\right)^{1/2}\exp\left(-\frac{\rho'^2}{2w^2_0}\right)\left(\frac{\rho'}{w_0}\right)^{|\ell'|}
L^{|\ell'|}_{p'}\left(\frac{\rho'^2}{w^2_0}\right)e^{i\ell'\phi'} \, .
\end{equation}
Our expansion has the form
\begin{equation}
\label{Eq30}
\psi(x',y') = \sum_{\ell'=-\infty}^\infty \sum_{p'=0}^\infty c_{p'\ell'} u_{p'\ell'}(\rho',\phi')
\end{equation}
and we can exploit the orthonormality of the Laguerre-Gaussian modes to show
that the expansion coefficients have the form \cite{RobertaRes}:
\begin{equation}
\label{Eq31}
c_{p'\ell'} = (-1)^{p'}\left(\frac{1}{(|\ell'|+p')!p'!}\right)^{1/2}\left(-\frac{x_0}{2w_0}\right)^{2p'+|\ell'|}\exp\left(-\frac{x^2_0}{4w^2_0}\right) \, .
\end{equation}
It is clear, in particular, that modes with $\ell' = +1$ are present in the
superposition.  It is from these modes that the atom can absorb a single
photon, acquiring in the process $+\hbar$ from the orbital angular momentum to
add to the single quantum of spin angular momentum.

We can use the expansion to determine the fraction of the total intensity that
is in modes with the required optical angular momentum to drive the quadrupole
transition.  It is convenient to express this in terms of the probability that
any single photon in the beam has orbital angular momentum $\hbar$:
\begin{align}
\label{Eq32}
P(\ell') &= \sum_{p'=0}^\infty |c_{p'\ell'}|^2 \nonumber \\
&= \exp\left(-\frac{x^2_0}{2w^2_0}\right)I_{\ell'}\left(\frac{x^2_0}{2w^2_0}\right) \, ,
\end{align}
where $I_n$ is the modified Bessel function of the first kind of order $n$.
It is clear that the mean value of $\ell'$ is zero.  This follows directly from
the fact that $I_{-n} = I_n$.  We can show by direct calculation from this
probability distribution, moreover, that
\begin{equation}
\label{Eq33}
\Delta\ell'^2 = \frac{x^2_0}{2w^2_0} \, ,
\end{equation}
as anticipated above.

\begin{figure}[h!]
\centering
\includegraphics[width=0.75\textwidth]{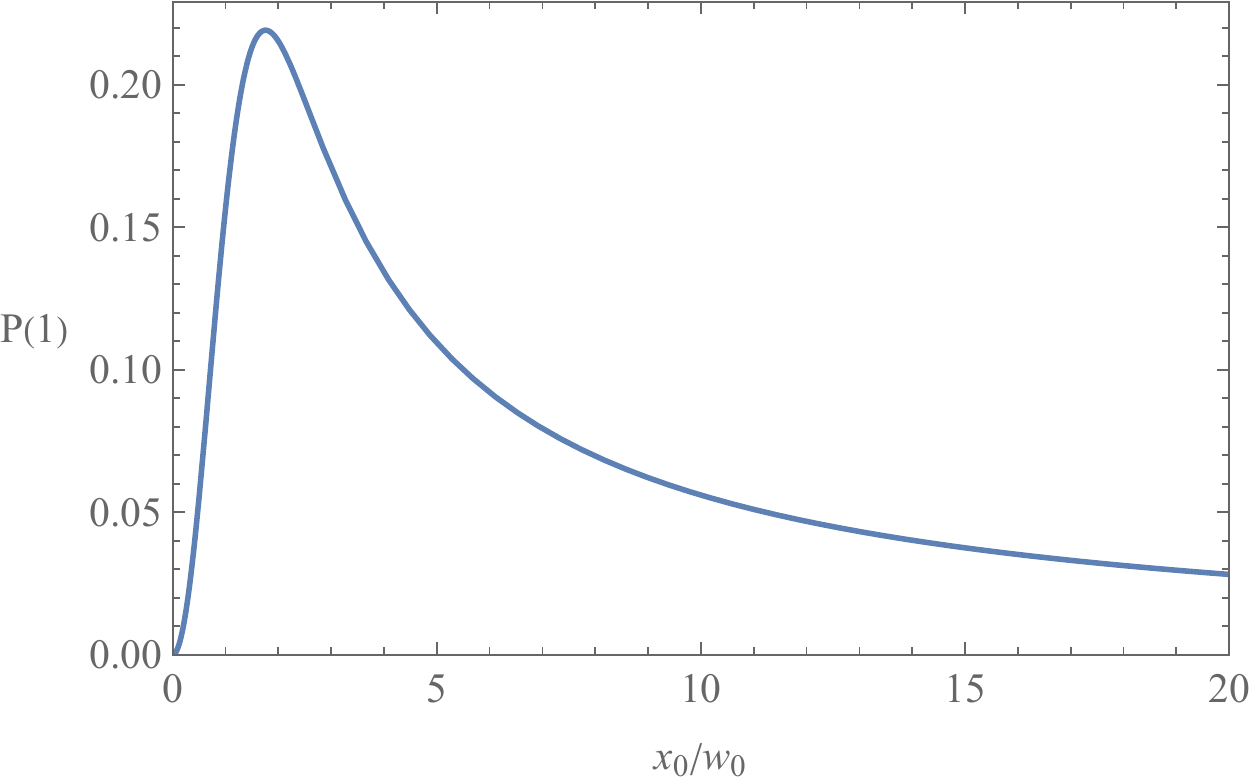}
\caption{The probability $P(1)$ that any single photon in the beam has
orbital angular momentum $\hbar$, as a function of relative position from the
beam waist $\frac{x_0}{w_0}$, as given by Eq.~\ref{Eq32}.}
\label{p_l_plot}
\end{figure}

The fraction of the light that is in the correct mode for driving our
quadrupole transition is $\exp(-x^2_0/2w^2_0)I_1(x^2_0/2w^2_0)$.
This probability is depicted in Figure~\ref{p_l_plot}.  We note that if the atom is positioned
on the beam axis, so that $x_0 = 0$, this probability is zero and the field cannot drive the
quadrupole transition.  As we move the position of the atom away from the beam axis the
probability that the local orbital angular momentum is $+\hbar$ increases and with it the
probability of exciting the transition.  At large distances from the beam axis the spread in
the distribution of the orbital angular momentum increases and with this the components
of the field having $\ell = +1$ decrease in amplitude.

\section{Resolution: orbital angular momentum is quasi-intrinsic}

It is certainly true that the optical orbital angular momentum is an intrinsic
quantity in that the total orbital angular momentum is unchanged by a parallel
displacement of the axis about which it is calculated.  Yet this is not the
whole story as such a displacement leaves the total mean orbital angular
momentum unchanged, but the mode is not an eigenmode of the orbital angular
momentum about the displaced axis.  In this sense it is, perhaps, better to
think of the optical orbital angular momentum as a quasi-intrinsic quantity
\cite{RobertaPRL}: it is, in effect, intrinsic on average.

An atom positioned off the axis of a circularly polarised Gaussian beam can
interact with any of a superposition of eigenmodes of $L'_z$ (about an axis
passing through the atom).  This means that the orbital angular momentum
required to conserve angular momentum in a quadrupole transition is readily
available to the atom at every position except on the beam axis.


\section{Absorption from a beam with no angular momentum}

As a final point we ask how our quadrupole transition can be driven by a
Gaussian beam that is linearly polarised.  The issue here is that such a beam
carries zero orbital angular momentum, but also {\it zero} spin angular
momentum.  We have seen that an atom that is not on the beam axis sees a field
that is a superposition of orbital angular momentum modes about an axis passing
through the atom and, in this way, the field can provide the required quantum
of orbital angular momentum.

If the field is linearly polarised then it carries {\it zero} units of spin
angular momentum.  As the spin angular momentum is intrinsic, it is zero at
every point in a linearly polarised beam and we can ask where the required
quantum of spin angular momentum comes from.  The solution to this question
mirrors that for the orbital angular momentum paradox in that a linearly
polarised beam is an equally weighted superposition of left- and right-handed
circularly polarised beams.  This follows simply from the decomposition
\begin{equation}
\label{Eq34}
\hat{\bf x} = \frac{1}{\sqrt{2}}\left[\frac{1}{\sqrt{2}}\left(\hat{\bf x} + i\hat{\bf y}\right) + \frac{1}{\sqrt{2}}\left(\hat{\bf x} - i\hat{\bf y}\right)\right] \, .
\end{equation}
It follows that half of the light in a linearly polarised beam carries the
required spin angular momentum to drive either a $\Delta m = +1$ electric
dipole transition or, in combination with the correct component orbital angular
momentum mode, a $\Delta m = +2$ electric quadrupole transition.


\section{Conclusion}

The orbital angular momentum of a light beam in the direction of propagation is
intrinsic, which means that its value is the same everywhere in the beam
\cite{Berry}.  Our paradox was that given this intrinsic nature, where does the
required angular momentum come from to drive a $\Delta m = +2$ electric
quadrupole transition?  The resolution, inspired by quantum theory, is that
modes of well-defined orbital angular momentum about the beam axis do not also
have well-defined orbital angular momentum about a displaced axis; the orbital
angular momenta about the $z$- and $z'$-axes are, in effect, incompatible
quantities.

It follows that a Gaussian beam with precisely zero orbital angular momentum
per photon about the beam axis is also a superposition of modes with all
possible orbital angular momenta about any axis parallel to that of the beam
\cite{RobertaRes,RobertaPRL}. The orbital angular momentum required to drive an
electric quadrupole transition derives from the modes in this superposition
that have $+\hbar$ orbital angular momentum about an axis passing through the
atom.  This exists at all positions in the beam except on the beam axis.

The ideas and the paradox presented and resolved here apply, also, to higher
order multipolar transitions.  An electric octopole transition, with $\Delta m
= +3$ for example \cite{Roberts}, can be driven by a circularly polarised
Gaussian beam if the absorbing atom is away from the beam axis.  In this case
there are two quanta of orbital angular momentum required and these will come
from those modes in the superposition with $\ell' = +2$.


\acknowledgments

\noindent It is a pleasure to dedicate this work to Michael Berry who has for
so long been an inspiration to those who work in optics and in mathematical
physics.  SMB thanks the Royal Society for the award of a Research
Professorship (RP150122).




\section*{References}

\begin{thebibliography}{99}

\bibitem{Kepler} Kepler J 1619 {\it De cometis libelli tres} (Augustae Vindelicorum: A. Apergeri)

\bibitem{Maxwell} Maxwell J C 1998 {\it A treatise on electricity and magnetism} vol. 2
(Oxford: Oxford University Press) Art. 793

\bibitem{Poyntingvec}
Poynting J H 1884 On the transfer of energy in the electromagnetic field {\it Phil. Trans.}
{\bf 175} 343--361

\bibitem{Poyntingbook}
Poynting J H 1910 {\it The pressure of light} (London: Richard Clay and Sons)

\bibitem{Poyntingspin}
Poynting J H 1909 The wave-motion of a revolving shaft, and a suggestion as to the
angular momentum in a beam of circularly polarised light {\it Proc. Roy. Soc. A}
{\bf 82} 560--567

\bibitem{Les}
Allen L, Beijersbergen M W, Spreeuw R J C and Woerdman J P (1992) Orbital angular
momentum of light and the transformation of Laguerre-Gaussian laser modes {\it Phys Rev A}
{\bf 45} 8185--8189

\bibitem{OAMbook} Allen L, Barnett S M and Padgett M J 2003 {\it Optical
    angular momentum} (London: Institute of Physics Publishing)

\bibitem{Bekshaev} Bekhshaev A, Soskin M and Vasnetsov M 2008 {\it Paraxial
    light beams with orbital angular momentum} (New York: Nova)

\bibitem{Alison} Yao A M and Padgett M J 2011 Orbital angular momentum:
    origins, behavior and applications {\it Adv. Opt. Photon.} {\bf 3} 161--204

\bibitem{Andrews} Andrews D L and Babiker M eds. 2012 {\it The angular momentum
    of light} (Cambridge: Cambridge University Press)

\bibitem{PhilTrans} Barnett S M, Babiker M and Padgett M J eds. 2017 {\it Optical
orbital angular momentum} {\it Phil. Trans. R. Soc.} {\bf 375} issue 2087

\bibitem{Condon}
Condon E U and Shortley G H 1953 {\it The theory of atomic spectra}
(Cambridge: Cambridge University Press)

\bibitem{Kuhn}
Kuhn H G 1962 {\it Atomic spectra} (Bristol: Arrowsmith)

\bibitem{CohenT}
Cohen-Tannoudji C, Diu B and Lalo\"{e} F {\it Quantum mechanics} vol. 2
(New York: Wiley)

\bibitem{Sobelman}
Sobelman I I  1979 {\it Atomic spectra and radiative transitions} (Berlin: Springer-Verlag)

\bibitem{Rose}
Rose M E 1955 {\it Multipole fields} (New York: Wiley)

\bibitem{Bennett} Babiker M, Bennett C R, Andrews D L and D\'{a}vila Romero L C
    2002 Orbital angular momentum exchange in the interaction of twisted light
    with molecules {\it Phys. Rev. Lett.} {\bf 89} 143601

\bibitem{Lembessis}
Babiker M, Andrews D L and Lembessis V E 2019 Atoms in complex twisted light {\it J. Opt.} {\bf 21} 013001

\bibitem{BerryDennis}
Berry M V and Dennis M R 2004 Quantum cores of optical phase singularities {\it J. Opt. A: Pure Appl. Opt.}
{\bf 6} S178--180

\bibitem{Qcore}
Barnett S M 2008 On the quantum core of an optical vortex {\it J. Mod. Opt.} {\bf 55} 2279--2292

\bibitem{superkick}
Barnett S M and Berry M V 2013 Superweak momentum transfer near optical vortices {\it J. Opt.} {\bf 15}
125701

\bibitem{Kien}
Kien F L, Ray T, Nieddu T, Busch T and Chormaic S N 2018 Enhancement of the quadrupole interaction fo an atom
with the guided light of an ultrathin optical fiber {\it Phys. Rev. A} {\bf 97} 013821

\bibitem{Afanasev}
Afanasev A, Carlson C E, Schmiegelov T, Schultz J, Scmidt-Kaler F ad Solyanik M 2018 Experimental verification of
position-dependent angular-momentum selection rules for absorption of twisted light by a bound electron {\it New J.
Phys.} {\bf 20} 023032

\bibitem{Darwin}
Darwin C G 1931 Notes on the theory of radiation {\it Proc. R. Soc. Lond. A} {\bf 136} 36--52

\bibitem{vanEnk}
van Enk S J and Nienhuis G 1994 Spin and orbital angular momentum of photons {\it Europhys. Lett.}
{\bf 25} 497--501

\bibitem{Rot}
Barnett S M 2010 Rotation of electromagnetic fields and the nature of optical angular momentum {\it J. Mod. Opt.}
{\bf 57} 1339--1343

\bibitem{natures}
Barnett S M, Allen L, Cameron R P, Gilson C R, Padgett M J, Speirits F C and Yao A M 2016 On the natures of the
spin and orbital parts of optical angular momentum {\it J. Opt.} {\bf 18} 064004

\bibitem{Anna}
O'Neil A T, MacVicar I, Allen L and Padgett M J 2002 Intrinsic and extrinsic nature of the orbital angular momentum
of light {\it Phys. Rev. Lett.} {\bf 88} 1--4

\bibitem{Berry}
Berry M V 1998 Paraxial beams of spinning light {in Singular optics} eds. Soskin M S and Vasnetsov M V SPIE 3487
6--11

\bibitem{Edwin}
Power E A 1964 {\it Introductory quantum electrodynamics} (London: Longmans)

\bibitem{Rodney}
Loudon R 2000 {\it The quantum theory of light} 3rd ed. (Oxford: Oxford University Press)

\bibitem{RoseAM}
Rose M E 1995 {\it Elementary theory of angular momentum} (New York: Dover)

\bibitem{Brink}
Brink D M and Satchler G R 1968 {\it Angular momentum} 2nd ed. (Oxford: Oxford University Press)

\bibitem{Edmonds}
Edmonds A R 1974 {\it Angular momentum in quantum mechanics} (Princeton University Press: Princeton NJ)

\bibitem{BiedenharnEd}
Biedenharn L C and van Dam H eds. 1965 {\it Quantum theory of angular momentum} (New York: Academic Press)

\bibitem{BiedenharnBook}
Biedenharn L C and Louck J D 1985 {\it Angular momentum in quantum physics} (Cambridge: Cambridge University Press)

\bibitem{Zare}
Zare R N 1988 {\it Angular momentum: understanding spatial effects in chemistry and physics} (New York: Wiley)

\bibitem{Greenshields}
Greenshields C R, Franke-Arnold S and Stamps R L 2015 Parallel axis theorem for free-space electron wavefunctions
{\it New J. Phys.} {\bf 17} 093015

\bibitem{Kibble}
Kibble T W B 1973 {\it Classical mechanics} 2nd ed. (London: McGraw-Hill)

\bibitem{Siegman}
Siegman A E 1986 {\it Lasers} (Sausalito CA: University Science Books)

\bibitem{Milonni} Milonni P W and Eberly J H 2010 {\it Laser Physics} (Hoboken NJ: Wiley)

\bibitem{RobertaRes} Barnett S M and Zambrini R 2006 Resolution in rotation
    measurements {\it J. Mod. Opt.} {\bf 53} 613--625

\bibitem{RobertaPRL}
Zambrini R and Barnett S M 2006 Quasi-intrinsic angular momentum and the measurement of its spectrum
{\it Phys. Rev. Lett.} {\bf 96} 113901

\bibitem{Roberts}
Roberts M, Taylor P, Barwood G P, Gill P, Klein H A and Rowley W R C 1997 Observation of an electric octopole
transition in a single ion {\it Phys. Rev. Lett} {\bf 78} 1876




\end{thebibliography}
\end{document}